\documentclass[showpacs,preprintnumbers,amsmath,amssymb,twocolumn,nofootinbib,prd]{revtex4-1}
\usepackage{graphicx}
\usepackage{slashed}

\usepackage[colorlinks,citecolor=blue,urlcolor=blue,linkcolor=blue]{hyperref}

\begin{document}

\title{Probing bino-wino coannihilation dark matter under the neutrino floor at the LHC}

\author{Guang Hua Duan$^{1,2,3}$}
\author{Ken-ichi Hikasa$^{4}$}
\author{Jie Ren$^{2,3}$}
\author{Lei Wu$^{1}$}
\author{Jin Min Yang$^{2,3,4}$}

\affiliation{
$^1$ Department of Physics and Institute of Theoretical Physics, Nanjing Normal University, Nanjing, Jiangsu 210023, China\\
$^2$ CAS Key Laboratory of Theoretical Physics, Institute of Theoretical Physics, Chinese Academy of Sciences, Beijing 100190, China\\
$^3$ School of Physical Sciences, University of Chinese Academy of Sciences, Beijing 100049, China\\
$^4$ Department of Physics, Tohoku University, Sendai 980-8578, Japan
}

\begin{abstract}
In the minimal supersymmetric standard model (MSSM) the bino-wino coannihilation
provides a feasible way to accommodate the observed cosmological dark matter (DM)
relic density. However, such a scenario usually predicts a very small DM-nucleon
scattering cross section that is below the neutrino floor, and can not be tested
by DM direct detection experiments. In this work, we investigate the discovery
potential of this bino-wino co-annihilation region by searching for
the soft dilepton events from the process
$pp \to \chi^0_2 (\to \ell^+\ell^- \chi^0_1) \chi^\pm_1+jets$ at the LHC.
We find that the mass of the wino-like $\chi^0_2$ can be probed up to about
310 (230) GeV at $2\sigma$ ($5\sigma$) level for an integrated luminosity
${\cal L}=300$ fb$^{-1}$. In the future HL-LHC with 3000 fb$^{-1}$ luminosity,
the corresponding mass limits can be pushed up to 430 (330) GeV.
\end{abstract}


\maketitle


\section{Introduction}
Dark matter accounts for about 27\% of the global energy budget in the Universe. However, it has
not yet been directly detected and its identity remains a mystery. The paradigm of weakly
interacting massive particles (WIMPs) is one of the most compelling versions and motivates
various searches for WIMP DM in the underground, collider and satellite
experiments~\cite{Arcadi:2017kky}.

Recently, the direct detections of nucleon-WIMP DM scattering have reached impressive
sensitivities. Their null results excluded the spin-independent cross section above
$7.7 \times 10^{-47}$ cm$^2$ for a DM mass of 35 GeV~\cite{xenon-1t} and the spin-dependent
cross section above $4.1 \times 10^{-41}$ cm$^2$ for a DM mass of 40 GeV~\cite{pandaX},
which are approaching the neutrino floor. Below this irreducible neutrino background,
the coherent neutrino scattering will mimic the DM signal. On the other hand, the WIMP
DM may interact with the nucleons at loop levels~\cite{Fox:2008kb} or annihilate with
a light species that is nearby in mass~\cite{Griest:1990kh}. Then, the correlation between
the DM annihilation cross section for relic density and the DM-nucleon scattering cross
section for direct detection will become weak or even disappear. Such a kind of WIMP DM
can therefore successfully meet the requirement of the observed relic density and escape
the current direct detections.

The minimal supersymmetric standard model (MSSM) is one of the most promising extensions
of the SM, in which the lightest neutralino ($\chi_1^0$) can be a natural WIMP dark matter
if $R$-parity is conserved. Provided that $\chi_1^0$ is bino-like\footnote{The lightest neutralino $\chi^0_1$ is
a linear combination of the gauge-eigenstates $(\tilde{B}, \tilde{W}, \tilde{H}_d^0, \tilde{H}_u^0)$: $\chi^0_1=N_{11}\tilde{B}+N_{12}\tilde{W}+N_{13}\tilde{H}_d^0+N_{14}\tilde{H}_u^0$, where $N$ is a unitary $4\times 4$ matrices that diagonalizes the neutralino mass matrix in Eq.~\ref{mass}. When $N_{11}^2 > \mathrm{max}\{N_{12}^2,N_{13}^2+N_{14}^2\}$, $\chi^0_1$ is called bino-like.},
there are usually three ways to achieve the correct DM relic density:
\begin{itemize}
 \item $\chi_1^0$ may be a well-tempered neutralino \cite{ArkaniHamed:2006mb}, i.e. an
appropriate mixture of bino and higgsino (or bino, higgsino and wino). However, a rather large
portion of the well-tempered bino-higgsino scenario is in tension with current XENON1T and LUX
data~\cite{Badziak:2017the,Abdughani:2017dqs,Baer:2016ucr}, and the entire parameter space could
be covered by future direct detections, such as LUX-ZEPLIN~\cite{LZ}. Even for the ``blind spots''~\cite{blindspot}, they may be unblinded with the collective efforts in future DM searches~\cite{Han:2016qtc};
 \item $\chi_1^0$ can annihilate resonantly via $Z$-funnel, Higgs-funnel or $H/A$-funnel, if its
mass is close to half of the mass of the funnel-particle. These funnel regions will be likely
covered by future (in)direct detections;
 \item $\chi_1^0$ can coannihilate with other light sparticles (e.g. a stop~\cite{Boehm:1999bj},
stau~\cite{Ellis:1998kh}, wino~\cite{Baer:2005jq}, or gluino~\cite{Profumo:2004wk}). This
requires their masses to be nearly degenerate. Due to the sterility of bino-like DM, the
DM-nucleon scattering rates in these co-annihilation scenarios are usually below the neutrino
floor.
\end{itemize}
Given the strong LHC constraints on the colored SUSY particles and the non-observation of DM in
direct detections, the bino-wino coannihilation is one of the feasible ways that can provide
the correct DM relic density and escape the direct
detections~\cite{Baer:2005jq,Ibe:2013pua,Profumo:2017ntc}. It may happen in the so-called split
supersymmetry~\cite{ArkaniHamed:2004fb,Giudice:2004tc,ArkaniHamed:2004yi,Wells:2004di},
the spread supersymmetry~\cite{Hall:2011jd} or supersymmetric GUT models with non-universal
gaugino masses at the boundary~\cite{Baer:2005jq}. Besides, such a scenario is also favored by
the very recent likelihood analysis of the pMSSM11 under various experimental
constraints~\cite{Bagnaschi:2017tru}.

In this work, we will investigate the potential of probing such a bino-wino coannihilation
scenario at the LHC. The ATLAS and CMS collaborations have widely searched for the
electroweakinos in different final states~\cite{Aaboud:2018jiw,Sirunyan:2018ubx}. However,
the sensitivity of these channels depends on the mass difference ($\Delta m$) between NLSP
and LSP. When $\Delta m$ becomes small, the productions of electroweakinos would lead to a
small missing energy and soft leptons in the final states,
even a displaced vertex~\cite{Rolbiecki:2015gsa,Nagata:2015pra} at the LHC.
In order to detect those compressed electroweakinos with the mass splitting $\Delta m < 5$ GeV,
one often uses a visible object $X$ ($X=j,\gamma,Z$) from initial state radiation (ISR) to boost
the electroweakino pair system and enhance $\slashed E_T$ in the event, while the other decay
products still remain soft~\cite{Han:2013usa,Schwaller:2013baa,Anandakrishnan:2014exa,Baer:2014cua,Low:2014cba,Barducci:2015ffa}. For the mass splitting $5~{\rm GeV}<\Delta m< 50$ GeV, the low
transverse momentum lepton events from the decays of compressed electroweakinos can be used
to effectively distinguish the signal from the SM
backgrounds~\cite{Han:2014kaa,Bramante:2014dza,Han:2014xoa,Baer:2014kya,Han:2015lma}. We will utilize such
soft dilepton events from the process $pp \to \chi^0_2 (\to \ell^+\ell^- \chi^0_1) \chi^\pm_1 + jets$ to probe the bino-wino coannihilation
at the LHC. Besides, there is no upper limit on the number of jets in our analysis. Although a veto on the second and other jets with $p_T>20$ GeV in the monojet analysis can sufficiently suppress the $t\bar{t}$ background, we will lose too many signal events~\cite{Drees:2012dd}. Instead, one can apply a cut on the transverse mass $m_{T}(\ell_i,\slashed E_T)$ to be below 70 GeV, which can also remove the $t\bar{t}$ background events efficiently but do not hurt the signal too much. Then, being inclusive in jets can increase the signal events and give us better sensitivity than the conventional monojet method.

The structure of this paper is organized as follows. In Section~\ref{section2}, we confront the bino-wino coannihilation scenario with the direct detections. In Section~\ref{section3}, we present the discovery potential of this scenario by analyzing the soft dilepton events at the LHC. Finally, we draw our conclusions in Section~\ref{section4}.

\section{bino-wino coannihilation and direct detections}
\label{section2}
In the MSSM the mass matrix for neutralinos $\chi_{1,2,3,4}^0$ in gauge-eigenstate
basis $(\tilde{B}, \tilde{W}, \tilde{H}_d^0, \tilde{H}_u^0)$ is given by
  \footnotesize
  \begin{eqnarray}
  \label{mass}
  M_{\chi^0}=
\left(
   \begin{array}{cccc}
     M_1 &0&-c_{\beta}s_Wm_Z&s_{\beta}s_Wm_Z\\
     0& M_2&c_{\beta}c_Wm_Z&s_{\beta}c_Wm_Z\\
     -c_{\beta}s_Wm_Z&c_{\beta}c_Wm_Z&0&-\mu\\
     s_{\beta}s_Wm_Z&-s_{\beta}c_Wm_Z&-\mu&0\\
     \end{array}
 \right)
\end{eqnarray}
 \normalsize
where $s_\beta = \sin\beta$ and $c_\beta = \cos\beta$. $M_{1,2}$ are soft-breaking mass parameters for
bino and wino, and $\mu$ is the higgsino mass parameter. This mass matrix can be diagonalized by a
unitary $4\times 4$ matrices $N$~\cite{matrix}. The chargino mass matrix in the gauge-eigenstates
basis ($\tilde{W}^+, \tilde{H}_u^+$, $\tilde{W}^-, \tilde{H}_d^-$) is given by
 \begin{eqnarray}
  \mathrm{M}_{\chi^{\pm}}=
\left(
   \begin{array}{cc}
     0 &X^T\\
     X& 0\\
     \end{array}
 \right)
\end{eqnarray}
with
\begin{eqnarray}
  X=
\left(
   \begin{array}{cc}
     M_2 &\sqrt{2}s_{\beta}m_W\\
     \sqrt{2}c_{\beta}m_W& \mu\\
     \end{array}
 \right)
\end{eqnarray}
Here the mass matrix $X$ can be diagonalized by two unitary $2\times 2$ matrices $U$ and $V$~\cite{matrix}.

In the limit of $|M_1| \simeq |M_2|\ll \mu$, one can have the masses of $\chi^0_{1,2}$ and $\chi^\pm_1$ at ${\cal O}(1/\mu)$:
\begin{eqnarray}
   m_{\chi_1^0} &\simeq& |M_1|-\frac{m_W^2\tan^2\theta_W}{\mu}\sin 2\beta, \\
   m_{\chi_2^0,\chi^\pm_1} &\simeq& |M_2|-\frac{m_W^2}{\mu}\sin 2\beta.
\end{eqnarray}
with the mass splitting
\begin{equation}
\Delta m_{\chi_2^0,\chi_1^\pm - \chi_1^0} \simeq \left(|M_2|-|M_1|\right) \\ -\frac{m_W^2\left(1-\tan^2\theta_W\right)}{\mu}\sin 2\beta.
\end{equation}
Due to the small couplings of Higgs/$Z$ boson with the bino-like $\chi^0_1$, the annihilation rate
of $\chi^0_1 \chi^0_1$ into the SM particles is highly suppressed. Then, the coannihilation of
$\chi^0_1$ with the wino-like electroweakinos $\chi^\pm_1, \chi^0_2$ will become essential
in producing the correct DM relic density.

We scan over the relevant parameter space for the bino-wino coannihilation and examine it under the DM direct detections:
\begin{eqnarray}
100~{\rm GeV}<|M_{1,2}|<1~{\rm TeV}, \quad 1< \tan\beta <60.
\end{eqnarray}
We assume a common mass $\tilde{M}$ for the higgsino mass parameter $\mu$, gluino mass parameter $M_3$, the pseudoscalar Higgs mass $m_A$, the first-two generation squark soft mass parameters $m_{\tilde{q}_{1,2}}$ and all slepton soft mass parameters $m_{\tilde{\ell}_{1,2,3}}$ and fix $\tilde{M}=5$ TeV for simplicity. In the MSSM with moderate $\tan \beta$ and large $m_A$, the Higgs mass is given by \cite{Carena:1995bx,Carena:1995wu}
\begin{eqnarray}\label{mh}
 m^2_{h}  \simeq M^2_Z\cos^2 2\beta +  \frac{3m^4_t}{4\pi^2v^2} \ln\frac{M^2_{S}}{m^2_t} \nonumber \\ +\frac{3m^4_t}{4\pi^2v^2}\frac{X^2_t}{M_{S}^2} \left( 1 -
\frac{X^2_t}{12M^2_{S}}\right),
\end{eqnarray}
\normalsize
where $v=174 {\rm ~GeV}$, $X_t \equiv A_t - \mu \cot \beta$ and $M_{S}$ is the average stop mass scale defined by $M_{S}=\sqrt{m_{\tilde{t}_1}m_{\tilde{t}_2}}$. From Eq.~\ref{mh}, it can be seen that the heavy multi-TeV stops and/or large Higgs-stop trilinear soft-breaking coupling are required to achieve a observed 125 GeV Higgs mass. We take the third generation squark soft masses $M_{\tilde{Q}_{3L}}=M_{\tilde{t}_{3R}}=M_{\tilde{b}_{3R}}=5$ TeV and vary the trilinear parameters in the range of $|A_{t,b}|<3$ TeV. We evaluate the mass spectrum and branching ratios of the sparticles with \textsf{SUSY-HIT} \cite{SUSYHIT} and impose the following constraints:
\begin{itemize}
\item[(1)] The light CP-even Higgs boson mass should be within the range of $125 \pm 3$ GeV.
\item[(2)] The DM relic density should satisfy the observed value $0.1186\pm 0.0020$ \cite{relic density} within $2\sigma$ range, which is computed by \textsf{MicrOMEGAs 4.3.2} \cite{micromega}.
\end{itemize}

\begin{figure}[ht]
  \centering
  \includegraphics[width=3.0in]{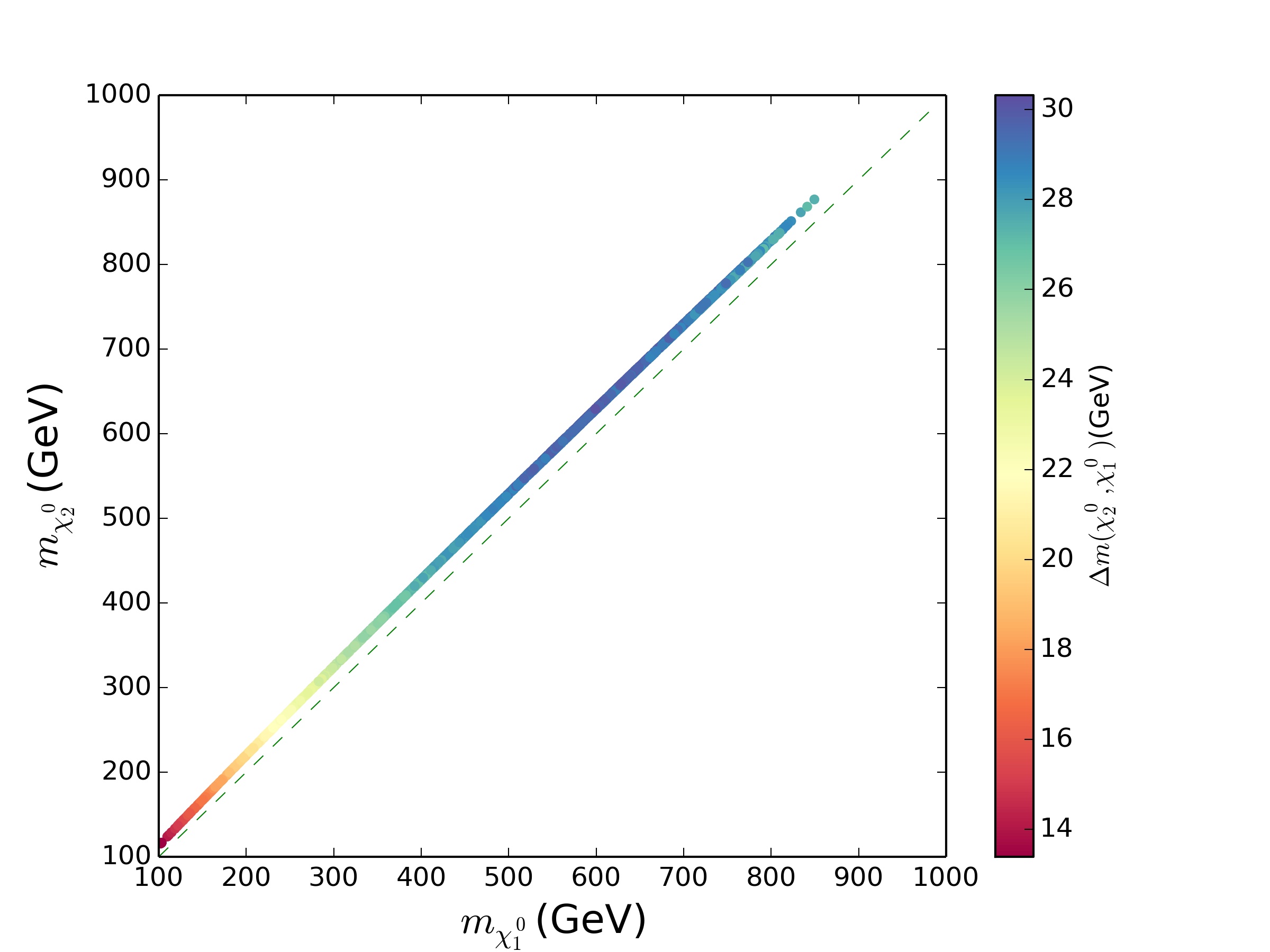}
  \vspace{-.3cm}
  \caption{The scatter plots of bino-wino coannihilation on the plane of $m_{\chi^0_1}$ versus $m_{\chi^0_2}$.
All samples satisfy the constraints (1) and (2).}
\label{n1n2}
\end{figure}
In Fig.~\ref{n1n2} we show the samples that satisfy the above constraints (1) and (2) on the plane of $m_{\chi^0_1}$ versus $m_{\chi^0_2}$. In order for the coannihilation partner $\chi^\pm_1$ to have a significant effect on the dark matter relic density, its mass should be close to $\chi^0_1$ with a splitting $\Delta m_{\chi^\pm_1 - \chi^0_1} / m_{\chi^0_1} \simeq 3\%-10\%$. The dominant contribution to achieving the relic density comes from $\chi^0_2\chi^\pm_1$ coannihilation ($\sim 50\%$), which is followed by $\chi^+_1\chi^-_1$ ($\sim 30\%$) and $\chi^0_1\chi^\pm_1/\chi^0_2/\chi^0_1$ ($\sim 20\%$) coannihilation processes. When $\Delta m_{\chi^\pm_1 - \chi^0_1} \ll M_W$, the wino-like chargino decays into a pair of fermions and a bino-like neutralino, via an off-shell $W$. The decay width of this process is given by,
\begin{equation}
\Gamma(\chi^\pm_1 \to f\bar{f}'\chi^0_1)\simeq \frac{4\alpha^2}{15\pi}\frac{(\Delta m_{\chi^\pm_1 - \chi^0_1})^3}{\mu^2}\frac{\sin^2{2\beta}}{\sin^2{2\theta_W}}.
\end{equation}
Very recently, ATLAS has published its updated analysis of searching for disappearing track events from the compressed chargino decay~\cite{Aaboud:2017mpt}, which excludes the chargino with lifetime longer than about 0.2 ns and masses less than 430 GeV. We checked that our samples can avoid such ATLAS limits because the lifetimes of our charginos are much shorter than 0.01 ns.

\begin{figure}[ht]
  \includegraphics[width=3in]{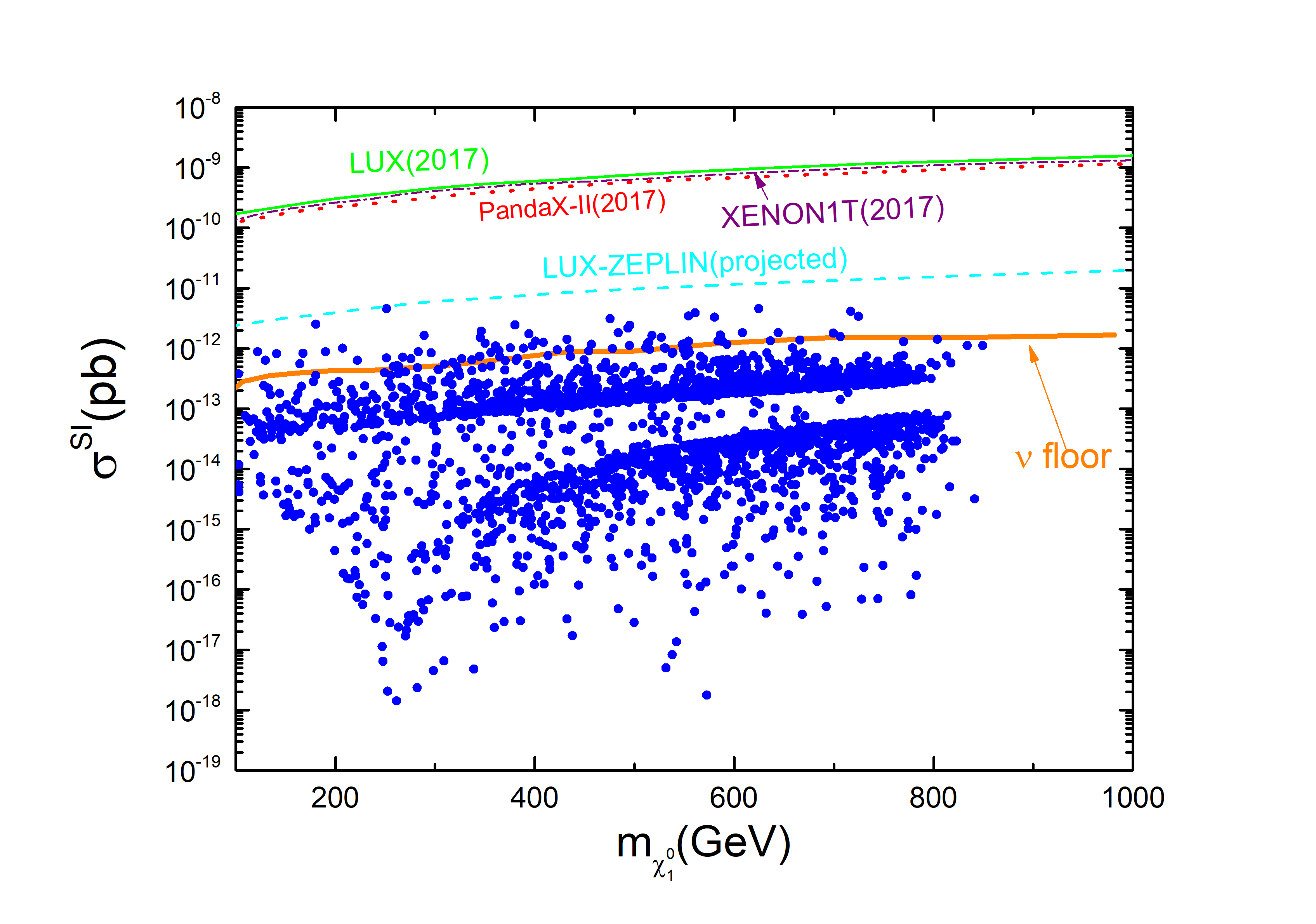}
  \includegraphics[width=3in]{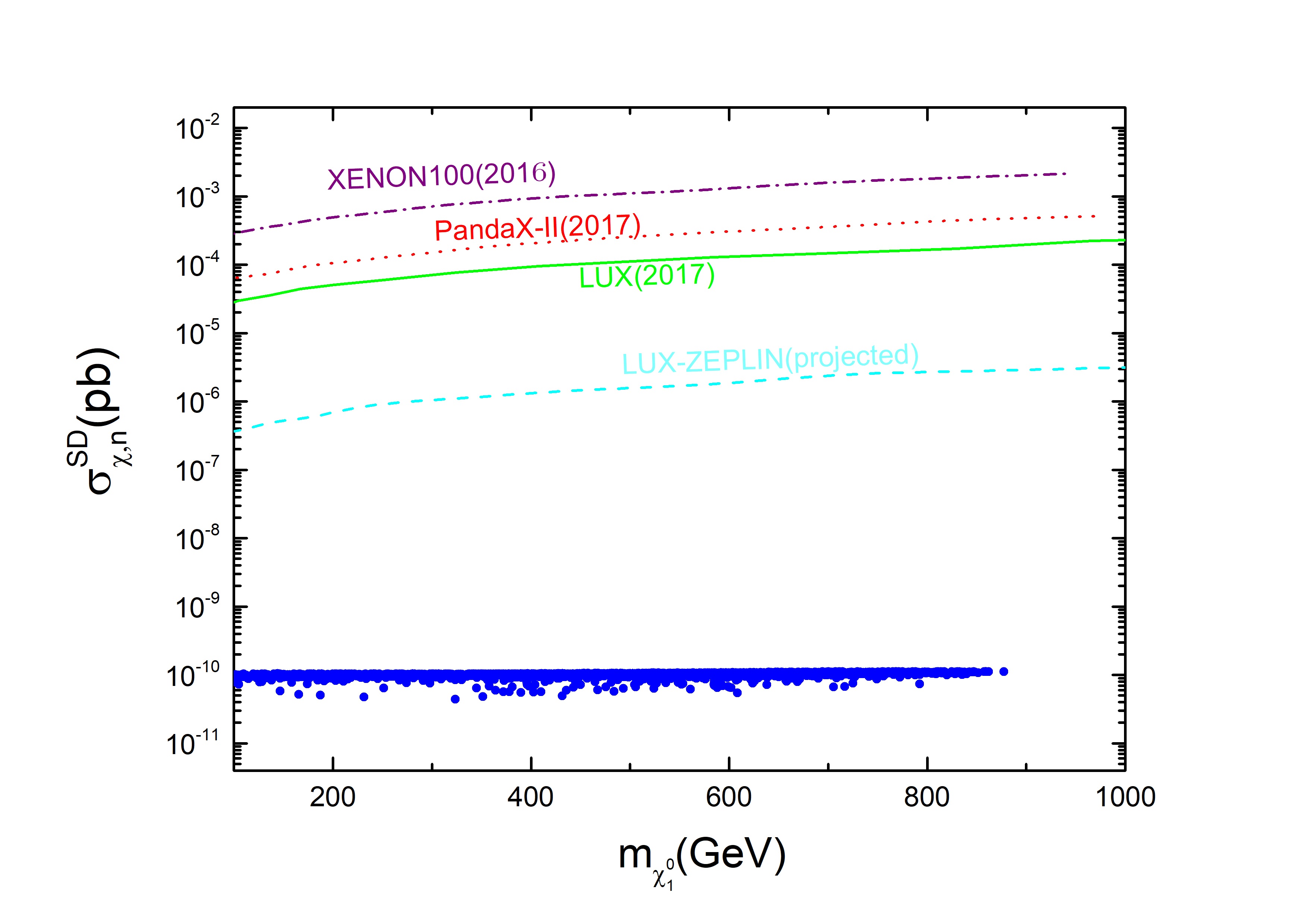}
  \caption{Same as Fig.~\ref{n1n2}, but for the spin-independent and spin-dependent neutralino LSP-nucleon scattering cross sections. The observed 90\% C.L. upper limits are from PandaX-II with a total exposure of 90.4 kg$\times$yr~\cite{pandaX}, XENON1T (2017) with a total exposure of 97.1 kg$\times$yr~\cite{xenon-1t}, XENON100 (2016) with a total exposure of 48 kg$\times$yr~\cite{Aprile:2016swn} and LUX (2017) with a total exposure of 129.5 kg$\times$yr~\cite{Akerib:2017kat}. The projected LUX-ZEPLIN with a exposure of $153.4 \times 10^2$ kg$\times$yr~\cite{Akerib:2018lyp} sensitivity limits are plotted.}
\label{dd}
\end{figure}
Next, we calculate the spin-independent (SI) and spin-dependent (SD) neutralino LSP-nucleon scattering cross sections, in which the form factors of proton and neutron are taken as,
\begin{eqnarray}
f^p_u \approx 0.020, \quad f^p_d \approx 0.026, \quad f^p_s \approx 0.13 \nonumber \\
f^n_u \approx 0.014, \quad f^n_d \approx 0.036, \quad f^n_s \approx 0.13
\end{eqnarray}
In Fig.~\ref{dd}, we show the direct detection results for all samples satisfying the constraints (1) and (2). We can see that the SI cross sections for bino-wino coannihilation samples are very small because of the tiny couplings of bino-like LSP with the Higgs boson~\footnote{In the top panel of Fig.~\ref{dd}, the separate bands for SI scattering cross section are caused by our scan method, in which we scan the parameter space for sign$(M_1/M_2)=\pm 1$ independently. The gap between two regions will become narrow and disappear when the number of samples for both sign$(M_1/M_2)=\pm 1$ increase enough.}. Most of them are even below the neutrino floor so that they can escape the existing strong limits from PandaX-II(2017), XENON1T(2017), XENON100(2016), LUX(2017) and the future direct detection experiments, such as LUX-ZEPLIN, which will improve the current sensitivity of LUX by about two orders of magnitude. Besides, it should be noted that the SD cross section of DM scattering off a neutron $\sigma^{SD}_{\chi,n}$ is largely determined by $Z$-boson exchange and is sensitive to the higgsino asymmetry, which can be approximately given by
\begin{equation}
\sigma^{SD}_{\chi,n} \simeq 3.1\times10^{-4}\text{pb} \cdot (\frac{|N_{13}|^2-|N_{14}|^2}{0.1})^2.
\end{equation}
The values of $|N_{13}|$ and $|N_{14}|$ for samples in our scan ranges are usually about ${\cal O} (10^{-2}) $ and ${\cal O} (10^{-3})$ so that $\sigma^{SD}_{\chi,n}$ is constantly around ${\cal O}(10^{-10})$ pb. Thus, it is meaningful to explore the
observability of this bino-wino coannihilation in other experiments, such as the LHC.

\section{Search for soft dilepton events at the LHC}\label{section3}
\begin{figure}[ht]
\includegraphics[width=2.5in]{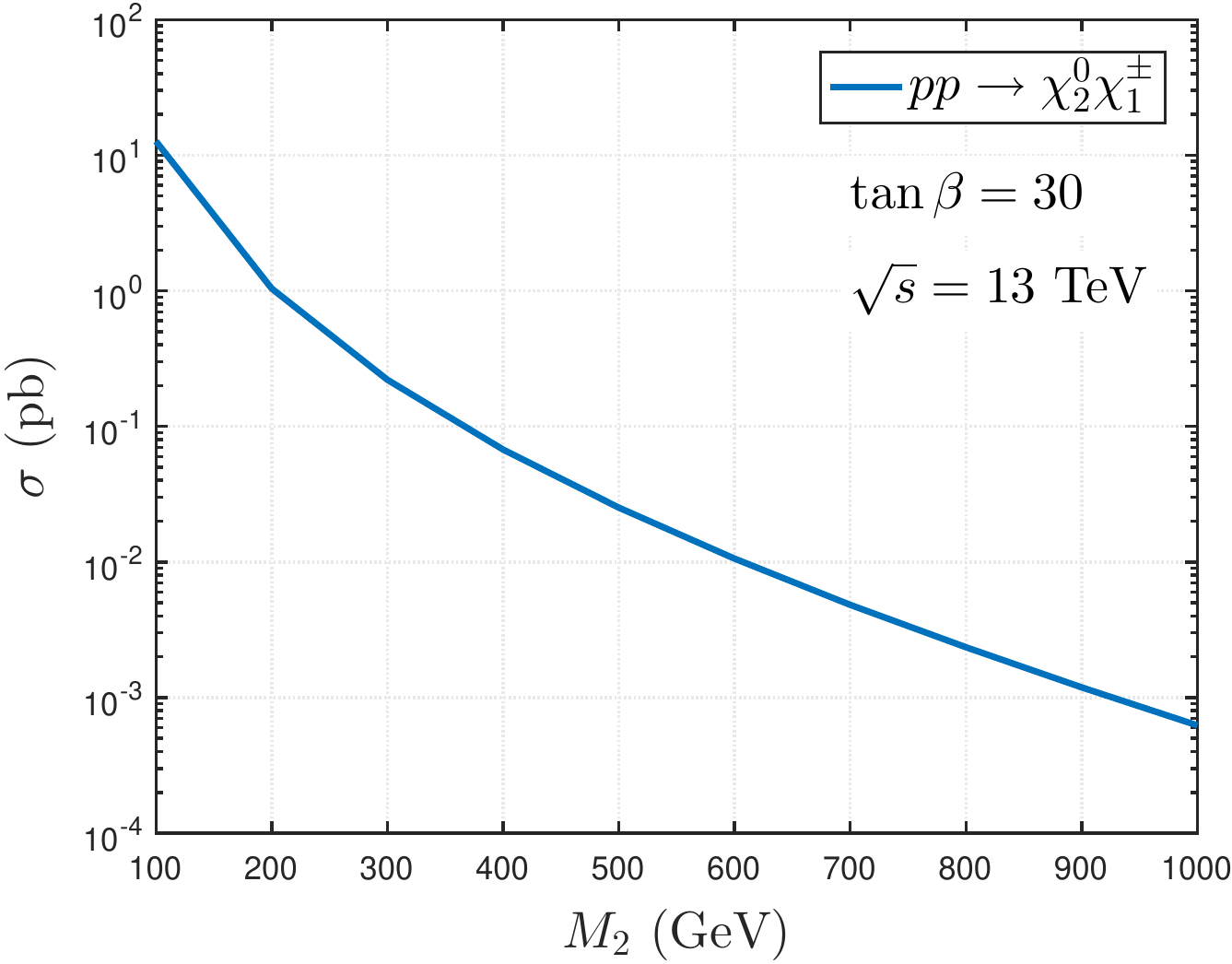}
\vspace{-0.3cm}
\caption{The cross section of $pp \to \chi^0_{2} \chi^\pm_1$ at the 13 TeV LHC.
Here $M_1=M_2-30$ GeV and $\tan\beta=30$.}
\label{cx}
\end{figure}
We calculate the cross section of the process $pp \to \chi^0_{2} \chi^\pm_1$
for $\tan\beta=30$ at 13 TeV LHC and show the result in Fig.~\ref{cx}.
We can see that the cross section can reach about 10 pb at $M_2=100$ GeV.
When $M_2$ becomes large, the cross section decreases rapidly.

\begin{figure}[ht]
  \centering
  \includegraphics[width=2.0in]{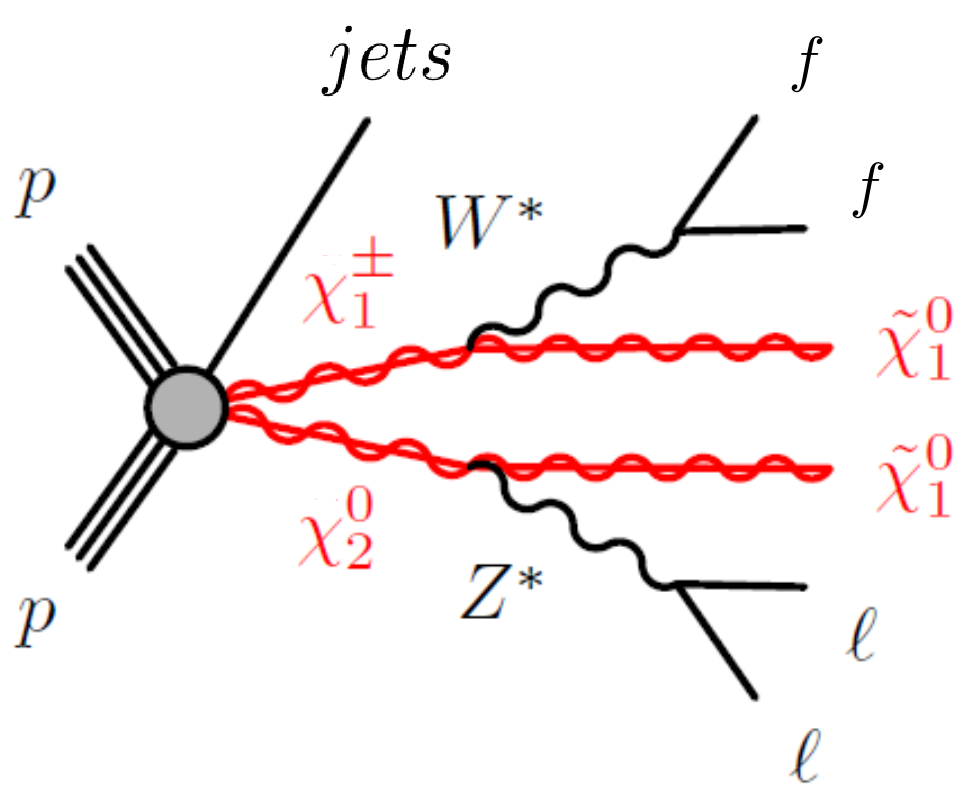}
  \vspace{-.3cm}
  \caption{Diagram representing for the process $pp \to \chi^0_2 \chi^\pm_1 + jets, \quad \chi^0_2 \to Z^*\chi^0_1 \to \ell^+\ell^- \chi^0_1$.}
\label{feyn}
\end{figure}
Now we perform a detailed Monte Carlo simulation of the process $pp \to \chi^0_2 \chi^\pm_1 + jets$ with the sequent decay channel $\chi^0_2 \to Z^*\chi^0_1 \to \ell^+\ell^- \chi^0_1$ at the LHC, as shown in Fig.~\ref{feyn}. Contrary to the monojet analysis for the compressed electroweakinos, there is no upper limit on the number of jets in our analysis. The main SM backgrounds for our signal are events from Drell-Yan (DY) processes with subsequent decays $\gamma/Z^* \to \tau^+\tau^- \to \ell^+\ell^- \nu_\ell \bar{\nu}_\ell \nu_\tau \bar{\nu}_\tau$,
and fully leptonic $t\bar{t}$ decays. Smaller backgrounds are diboson ($VV$) processes like $W^+W^-$
and single top production like $tW$. In additional, the non-prompt leptons can also mimic our signal,
which mainly arise from $W+j$ events. In Ref.~\cite{Curtin:2013zua}, several parameterizations of the
lepton fake rates are extracted from comparison with the CMS data. At low $p_T$, where the bulk of our
soft leptons distribute, the fake rate is quoted as ${\cal O}((0.6\hbox{--}3)\times 10^{-5})$. So, the background
from fake leptons is ${\cal O}(1 \%)$ of the dominant backgrounds. On the other hand, a low invariant
mass cut $4~{\rm GeV} < m_{\ell\ell} < 50~{\rm GeV}$ will further reduce such non-prompt backgrounds
as the distribution from $W+j$ events with a fake lepton has a higher value of $m_{\ell\ell}$ than
the signal~\cite{Han:2014kaa}. In our simulations, we focus on the prompt SM backgrounds.

\begin{table*}[t!]
\caption{The cut flow for the cross sections of the signal and backgrounds at the 13 TeV LHC
before using the signal regions. The benchmark point is $m_{\chi^0_1}=137.1$ GeV,
$m_{\chi^0_2}=m_{\chi^\pm_1}=153.3$ GeV, $\tan\beta=34$. The cross sections are in unit of fb.}
\label{cutflow}
\begin{tabular}{lp{1cm}rp{.5cm}rp{.5cm}rp{.5cm}r}
\hline\hline
&& \multicolumn{5}{c}{Backgrounds} && Signal \\
Cuts && $t\bar{t}$ && Drell-Yan && diboson && $\chi^0_2\chi^\pm_1$ \\
             \hline\hline
            $\slashed E_T > 125$ GeV && 45108.12 && 1636.39 && 2664.94 && 662.84\\
            \hline
            $N(j) > 0$, $N(b) = 0$ && 8776.16 && 1309.35 && 1903.41 && 528.24 \\
            \hline
            $N(\ell)=2$, OSSF, $p_T(\ell_{1,2}) \in [5,30]$ GeV && 57.45 && 31.45 && 5.77 && 5.45 \\
            \hline
             $m_{\ell\ell} \in [4, 50]$ GeV, $m_{\ell\ell} \notin [9, 10.5]$ GeV && 44.62 && 29.34 && 4.43 && 4.35 \\
            \hline
            $H_T >$ 100~GeV, $\slashed E_T / H_T \in [0.6, 1.4]$ && 28.14 && 23.47 && 3.32 && 3.59 \\
            \hline
            $M_T(l, \slashed E_T) < $ 70~GeV && 8.83 && 21.57 && 0.98 && 2.77 \\
            \hline
            $M_{\tau\tau} \notin [0, 160]$~GeV && 6.12 && 7.21 && 0.75 && 2.123 \\
            \hline\hline
\end{tabular}
\end{table*}

We generate the parton level events of signal and backgrounds with \textsf{MadGraph5\_aMC@NLO}~\cite{mad5} and implement the shower and hadronization by using \textsf{Pythia-8.2}~\cite{pythia}. The detector effects are simulated by \textsf{Delphes}~\cite{delphes}. The jet clustering is performed by \textsf{FastJet}~\cite{fastjet} with the anti-$k_t$ algorithm~\cite{anti-kt}. Due to the limitation of CPU, the signal, Drell-Yan and diboson production processes
are matched up to one jet by using the MLM-scheme with merging scale $Q=30$ GeV. We normalize the
cross sections of the processes $pp \to \chi^\pm_1 \chi^0_2$ and $pp \to t\bar{t}$ to their NLO order~\cite{prospino} and the NNLO+NNLL order values~\cite{Czakon:2011xx}, respectively.  The event selections are performed within the framework
of ~\textsf{CheckMATE2} \cite{checkmate}.

In our object reconstruction we require the leptons to satisfy $p_\mathrm{T} \in [5, 30]~\mathrm{GeV}$ and
$|\eta| < 2.5$. The leptons are isolated within a cone in $\eta$-$\phi$ space of radius
$\Delta R = \sqrt{\Delta\eta^2 + \Delta\phi^2} < 0.3$ if the sum of the transverse momenta
within the cone is less than 5 GeV and the ratio of it to the $p_\mathrm{T}$ of the lepton is
less than 0.5.  Jets are clustered from calorimeter towers using anti-$k_\mathrm{T}$ jet clustering
algorithm with a distance parameter $R = 0.4$. Jets are required to have $p_\mathrm{T} > 25~\mathrm{GeV}$
and $|\eta| < 2.4$. We assume the $b$-jet tagging efficiency as 80\% and include a misidentification
10\% for light jets.

The following cuts are applied to differentiate signal from backgrounds:
\begin{itemize}
  \item A large missing transverse energy $\slashed E_T > 125$ GeV is required.
  \item At least one jet but a veto on events with $p_T(b) > 25$ GeV is imposed to reduce the
 $t\bar{t}$ background.
  \item A pair of  opposite-sign same-flavor (OSSF) leptons are required. Two leptons should have
a small transverse momentum $5~{\rm GeV} <p_T(\ell_{1,2}) < 30~{\rm GeV}$.
  \item The invariant mass of the two leptons is required in the range
$4~{\rm GeV} <m_{\ell\ell}<50~{\rm GeV}$ but $m_{\ell\ell} \notin [9, 10.5]$ GeV, which can reduce
the diboson background and the potential soft lepton events from Drell-Yan and $J/\psi$ and $\Upsilon$
decays.
  \item The transverse hadronic energy $H_T>100$ GeV, which is defined as the scalar sum of the transverse momenta of the selected jets.
  \item The QCD multijet background can be
  efficiently suppressed by the requirement of large $\slashed E_T$ and two leptons. Besides, we apply a cut $0.6<\slashed E_T/ H_T<1.4$ to reject the residual QCD multijet events, which mostly sit at low $\slashed E_T/ H_T$ values~\cite{Sirunyan:2018iwl}. This is because that the large $\slashed E_T$ from the mismeasurement of QCD jets needs the mismeasured jets to have a large momentum as well, and therefore the $\slashed E_T/ H_T$ variable is quite small in QCD. In this respect, the lower cut on $\slashed E_T/ H_T$ is necessary, but not sufficient to suppress QCD. Together with the requirement of $\slashed E_T$ and two leptons, the QCD multijet background becomes negligible.
  \item The transverse mass $m_{T}(\ell_i,\slashed E_T)<70$ GeV is used to further suppress $t\bar{t}$ backgrounds.
  \item The invariant mass $m_{\tau\tau}\notin [0,160]$ GeV~\cite{Han:2014kaa}. This observable is used to reject the large background from $\gamma^* /Z \to \tau^+\tau^-$ events with the $\tau$ leptons decay leptonically. Since the $\tau$ leptons from $\gamma^*/Z$ boson decays are highly energetic, the direction of the final lepton is approximately the same as that of the parent $\tau$ lepton, which leads to $\vec{p}_{\nu_i}=\xi_i \vec{p}_{\ell_i}$. If the missing energy is due to four neutrinos, one can solve $\xi_i$ by using the measured $\slashed E_T$ and then reconstruct the $\tau$ lepton momentum through $p_{\tau_i}=p_{\nu_i}+p_{\ell_i}$. It should be noted that the $\tau$ lepton momentum magnitude can take a negative value when the flight direction is opposite to the lepton one. In those events, we set $m_{\tau\tau}$ to be negative. The $\gamma^* /Z \to \tau^+\tau^-$ backgrounds will have a narrow peak around $m_{Z}$ in $m_{\tau\tau}$ distribution, while the signal will be featureless.
\end{itemize}
After the above selections, we seperate the signal events into three regions:
$\slashed E_T \in [125,200]$ GeV, $\slashed  E_T \in [200,250]$ GeV,  $\slashed  E_T >250$ GeV.
In each $\slashed E_T$ bin, we further define four signal regions $m_{\ell\ell}=[4,10], [10,20], [20,30], [30,50]$ GeV to enhance the sensitivity.
In Table ~\ref{cutflow}, we give the cut flow for the cross sections of the signal and backgrounds
at the 13 TeV LHC before using the signal regions. As discussed above, we can see that the Drell-Yan
process is the largest background, which is followed by dileptonic $t\bar{t}$ process. The cut on the transverse mass $m_{T}(\ell_i,\slashed E_T)<70$ GeV can reduce the $t\bar{t}$ background heavily but do not hurt the signal too much.
The $m_{\tau\tau}$ cut plays an important role in suppressing the Drell-Yan background.

To estimate the observability, we evaluate the statistical significance ($\alpha$) by using Poisson
formula
\begin{eqnarray}\label{ss}
\alpha=\sqrt{2{\cal L}\left[(\sigma_S+\sigma_B){\rm ln}(1+\frac{\sigma_S}{\sigma_B})-\sigma_S\right]}.
\end{eqnarray}
where $\sigma_S$ and $\sigma_B$ are the signal and background cross sections and ${\cal L}$ is the
integrated luminosity.

\begin{figure}[ht!]
   \centering
  \includegraphics[width=3.5in,height=2.2in]{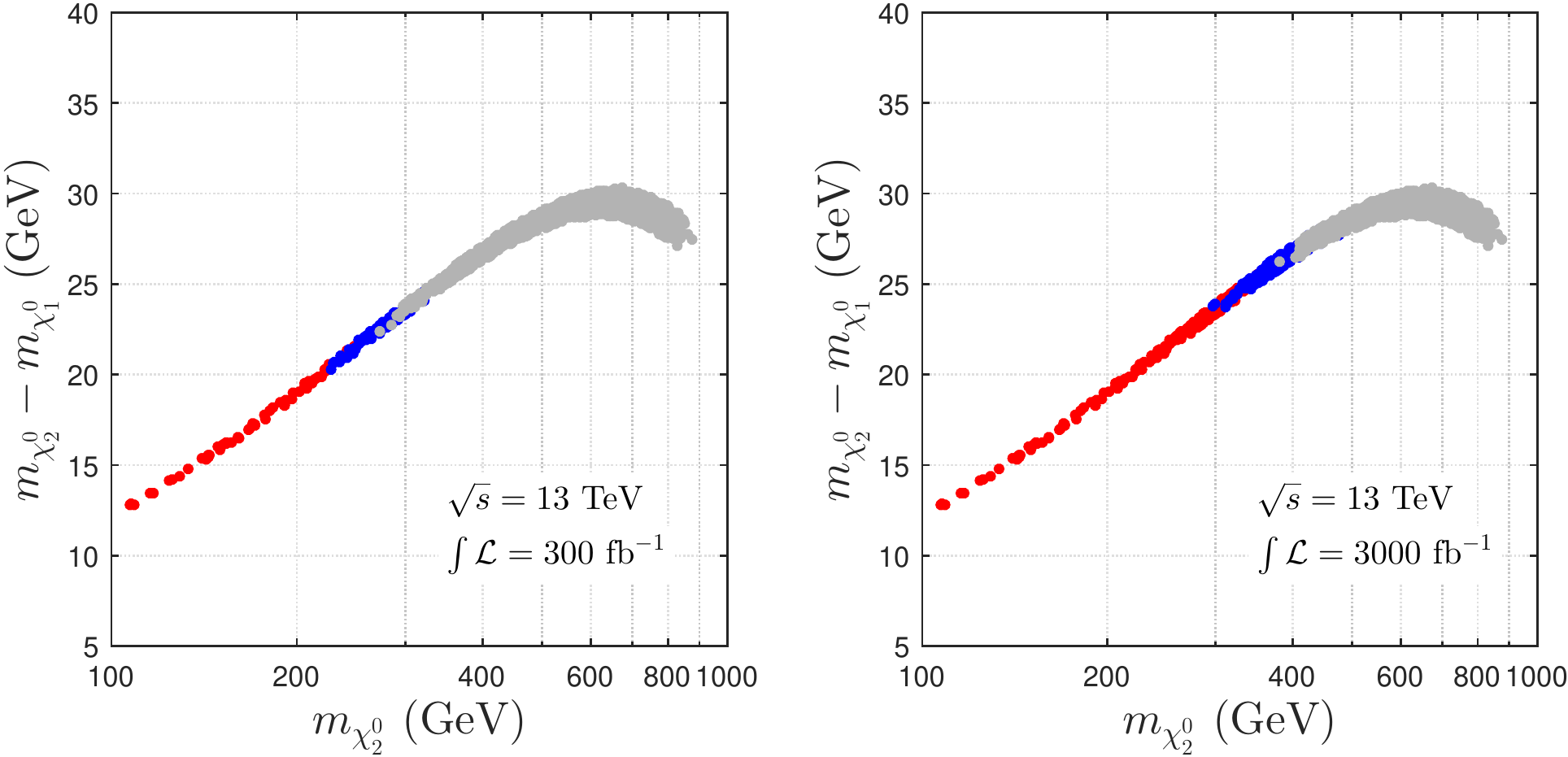}
  \caption{Same as Fig.~\ref{n1n2}, but show the statistical significance ($\alpha$) of the process
$pp \to \chi^0_2 \chi^\pm_1 + jets$ on the plane of $m_{\chi^0_2}$ versus $m_{\chi^0_2}-m_{\chi^0_1}$ at the 13 TeV LHC
with the luminosity ${\cal L}=300,~3000$ fb$^{-1}$. The gray, blue and red bullets denote
$\alpha <2\sigma$, $2\sigma< \alpha <5\sigma$ and $\alpha >5\sigma$, respectively.}
  \label{lhc}
\end{figure}
In Fig.~\ref{lhc} we present the statistical significance the process $pp \to \chi^0_2 \chi^\pm_1+jets$ for
the samples allowed by constraints (1-2) on the plane of $m_{\chi^0_2}$ versus $m_{\chi^0_2}-m_{\chi^0_1}$ at
the 13 TeV LHC. It can be seen that the significance decreases rapidly when the wino-like $\chi^0_2$
becomes heavy. The mass of $\chi^0_2$ can be probed up to about 310 (230) GeV at $2\sigma$ ($5\sigma$)
level for an integrated luminosity ${\cal L}=300$ fb$^{-1}$. In the future HL-LHC with 3000 fb$^{-1}$
luminosity, the corresponding mass limits can be pushed up to $m_{\chi^0_2}<$ 430 (330) GeV.

\section{conclusion}
\label{section4}
In the MSSM the bino-wino coannihilation is one of important ways to achieve the correct DM relic
density. Due to the small couplings of the bino-like DM with $Z/h$ bosons, such a scenario can not be
probed by the DM direct detections. In this work, we studied the discovery potential of the bino-wino
coannihilation scenario by searching for the soft dilepton from the three-body decay of the wino-like
neutralino $\chi^0_2$ in the process $pp \to \chi^0_2 \chi^\pm_1 + jets$ at the LHC. It turns out that the
wino-like $\chi^0_2$ can be probed for a mass up to about 310 (230) GeV at $2\sigma$ ($5\sigma$) level
for an integrated luminosity ${\cal L}=300$ fb$^{-1}$. The future HL-LHC with 3000 fb$^{-1}$ luminosity
will be able to push these mass limits up to 430 (330) GeV.

\section*{Acknowledgement}
We thank Mariana Frank for helpful discussions.
This work was supported by the National Natural Science Foundation of China
(NNSFC) under grant Nos. 11705093 and 11675242,
by Peng-Huan-Wu Theoretical Physics Innovation Center (11747601),
by the CAS Center for Excellence in Particle Physics (CCEPP),
by the CAS Key Research Program of Frontier Sciences and
by a Key R\&D Program of Ministry of Science and Technology of China
under number 2017YFA0402200-04.

\end{document}